# Testable Array Multipliers for a Better Utilization of C-Testability and Bijectivity

Fatemeh Sheikh Shoaei
*Electrical and Computer Engineering Department*
*University of Tehran*
Tehran, Iran
fns.shoaei@ut.ac.ir

Alireza Nahvy
*Electrical and Computer Engineering Department*
*University of Tehran*
Tehran, Iran
alireza_nahvy@ut.ac.ir

Zainalabedin Navabi
*Electrical and Computer Engineering Department*
*University of Tehran*
Tehran, Iran
navabi@ut.ac.ir

*Abstract*—This paper presents a design for test (DFT) architecture for fast and scalable testing of array multipliers (MULTs). Regardless of the MULT size, our proposed testable architecture, without major changes in the original architecture, requires only five test vectors. Test pattern generation (TPG) is done by combining C-testability, bijectivity and deterministic TPG methods. Experimental results show 100% fault coverage for single stuck-at faults. The proposed method requires minor testability hardware insertion into the multiplier with extra delay and area overhead of less than 0.5% for a 64-bit multiplier.

## I. Introduction

Currently, power and performance requirements have changed design trend focus from general-purpose to more application-specific cores. Therefore, many accelerator-rich architectures have recently been proposed [1-2]. Similarly, testing techniques need to migrate from general to more specific ones. Uusing test and testability methods customized for specific hardware, a.k.a. accelerator testing, area, power and performance overheads reduce and fault coverage enhances. An accelerator's computation part consumes most of the power and area while create timing bottleneck. Multipliers (MULTs), the core of accelerator's computation part, is vastly used in Deep Neural Networks (DNN), Finite Impulse Response (FIR) filters, Fast Fourier Transform (FFT) analyzers, feature extractors and other computer vision algorithm implementations.

MULT's vital role in such accelerators makes low overhead, testable MULT designs achieving high fault coverage be very crucial [3]. Although MULT testing has an old background, by the advent of adder and MULT-based accelerators (e.g., Coarse Grained Reconfigurable Architecture (CGRA), Tensor Processing Unit (TPU) [3] and pipeline MULT), this topic has received renewed attention. Traditional test methods such as scan-based testing due to their large area overhead are effective with large test vector count, while Built-In-Self-Test (BIST) is a better method with small test vector count.

Many computing accelerators have iterative, inter-connected cells. Architectures with such feature, a.k.a. Iterative Logic Arrays (ILAs), are efficient in terms of area (homogeneous placement and routing) and performance as well as design and test efforts [4]. One of the most practical hardware architectures for MULT core is array MULT [5], an ILA with iterative cells, each cell a combination of an AND gate and a Full Adder (FA).

The key to make an array multiplier testable can be seen in C-testability and bijectivity. An ILA is C-testable if it can be exhaustively tested with a constant number of test vectors independent of the array size [6]. Bijectivity of a single cell is an important condition of C-testability. A cell is bijective if its inputs and outputs are one-to-one mapped [6].

A bijective ILA can be tested pseudo-exhaustively [7] by applying the exhaustive tests only to the first cell. Shown in Fig. 1, each cell's response is the next cell's input test pattern. The last cell's response can be used for fault detection [4, 6]. Bijectivity provides high controllability and observability. A bijective ILA is controllable since correct selection of first cell test pattern makes cell$_{(i)}$'s test pattern controllable. Due to one-to-one mapping, each test vector has its unique response. Hence, a fault in every cell can be propagated into the final response, making this ILA observable [6].

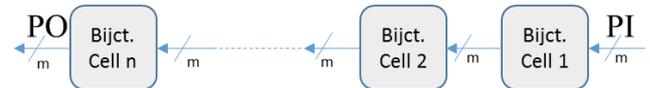

Fig. 1. Bijective ILA

This paper proposes a testable array MULT design that can be tested using five test vectors independent of the MULT size with 100% fault coverage, and low area, delay and power overheads. C-testability, bijectivity and deterministic test pattern generation (TPG) help achieve such scalability and testability.

This paper is organized as follows. Section II reviews related work. Section III discusses test methods for array MULT demanding no hardware modification. Section IV reviews adder's C-testability. Section V presents our proposed testable array MULT. Then, Section VI discusses how the proposed method combination with the cell-based deterministic TPG can reduce test vector count, while achieving 100% fault coverage. In section VII, a simple BIST architecture for our MULT testing is presented. Then, Section VIII and IX present experimental results and conclusions, respectively.

## II. RELATED WORKS

In the C-testability context, several works have been initiated. It was started in 1973 by Friedman [8] who worked theoretically on testing one-dimensional unilateral combinational ILAs. After that, several papers worked on different cases in ILA testing, like 2-dimensional ILAs (1967), bilateral communications between arrays (1981), sequential ILAs (1981), tree MULTs (1984), et al. [9]. The tree MULT was tested by 16 test vectors [10].

In 1990, Wu & Cappello [4], have used cells' bijectivity characteristics in several ILAs to reduce test vector count. In [11], an FFT is modified to become bijective, and a reconfiguration mechanism is proposed to bypass the faulty cell. Based on bijectivity, a test mechanism for Xilinx 7-series FPGAs has been proposed [12], recently.

Several works have focused on complete testing of MULTs (100% fault coverage) using C-testability. For example, in [13], a gate-level MULT is tested using 9 test vectors and another specific (not very common) MULT is tested using 6 test vectors. The first design has increased the critical path's delay due to the hardware modifications imposed on this path.

Reference [6] started with an adder and simply modified it by sending one of full adder's inputs to its output in order to make it bijective. Also, it added a multiplexer to each FA cell to improve its controllability. Using this idea, [6] made more complex designs bijective: MULT, MULT accumulator, carry save adder (CSA) MULT, FIR filter and matrix MULT.

In [6], cells have been combined to trade-off between hardware overhead and test vector count. Hardware overhead, associated with cell's testability insertion, is reduced by grouping cells and testability hardware sharing, although grouping demands larger test vectors for the combined cells. Such idea is used in [14] to modify a motion estimation system.

A combination of pseudo-exhaustive and deterministic testing methods based on C-testability is proposed for array MULTs [7]. This work imposes no design change and only requires MULT's row count minus one multiplied by 11 test vectors. This method's fault coverage is less than 100%. Reference [7] also proposes a BIST architecture. In [15], the same idea has been used. However, this method needs extra hardware to achieve 100% fault coverage and it considered sequential faults.

Our work uses a combination of C-testability, bijectivity and deterministic TPG to test an array MULT using only five test vectors, independent of the MULT size, with a negligible delay and power overhead. Also, our area overhead is low relative to the previous works.

## III. MULT TEST METHODS

Both traditional and novel MULT testing methods imposing no hardware overhead are discussed in this section.

### A. Random and Deterministic

Random testing, the most simple method, is neither efficient nor scalable for large and complex circuits [16]. Random testing is usually complemented by another test method (e.g., deterministic test generation) because its fault coverage cannot reach 100% without other methods [17]. As an example of deterministic test generation, Path-Oriented test generation (PODEM) is used to test a 4-bit MULT and 97.41% fault coverage is reached with 306 test vectors [18].

### B. C-Testability

Using C-testability for testing a design can help achieve 100% fault coverage with a small test vector count. The test vector count is constant regardless of the C-testable circuit size, making this technique fully scalable. However, C-testability needs specific properties not applicable to all circuits. However, some circuits can be modified to be C-testable.

Ripple Carry Adder (RCA) is a C-testable design intrinsically that can be tested with only eight test vectors for 100% fault coverage [8]. However, an array MULT is not C-testable intrinsically. In [7], C-testability is employed without any modifications to the MULT. It transformed a 2-dimensional array of array MULT to a set of one-dimensional arrays and used RCA C-testability feature on each one-dimensional array. The idea is to activate only one array MULT's row in each round and propagate the probable fault to the final row. The test vector count for a 4-bit MULT is 33, with less than 100% coverage since C-testability is partially applied to the MULT.

## IV. ADDER C-TESTABILITY

Testing a circuit using C-testability requires primary inputs to be selected such that all cells can be tested exhaustively. RCA is a popular completely C-testable circuit. Full Adders are RCA's cells chained by carry signals (Fig. 2).

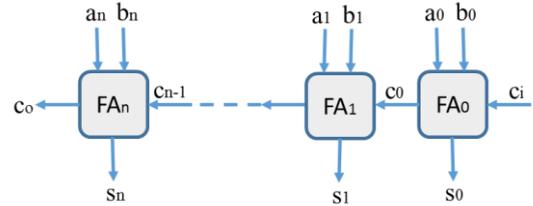

Fig. 2. Bijective ILA

Each cell (with three inputs) is tested by eight exhaustive test patterns as shown in Table I *(propagation table)*. Propagation table for a C-testable circuit lists all parallel inputs of a cell (A and B) and chain inputs ($C_{in}$) and outputs ($C_{out}$). The cell's $C_{in}$ is justified by back-tracing to the lower order chained cells *(row justification)*. Parallel inputs (A and B) justifying $C_{in}$'s desired value need to be propagated to the entire C-testable circuit *(row propagation)*. To justify a cell's chain input values, two test group exist. In one test group, the chain input is the same as the chain output (*simple row propagation*). In another test group, the complementary chain input is the same as the chain output (*complementary row propagation*). These definitions are used in TPG for the C-testable RCA shown in Fig. 2.

TABLE I. PROPAGATION TABLE FOR FA

| Num | Parallel inputs | | Chain input | Chain output |
|---|---|---|---|---|
| | *A* | *B* | *C-in* | *C-out* |
| 0 | 0 | 0 | 0 | 0 |
| 1 | 0 | 0 | 1 | 0 |
| 2 | 0 | 1 | 0 | 0 |
| 3 | 0 | 1 | 1 | 1 |
| 4 | 1 | 0 | 0 | 0 |
| 5 | 1 | 0 | 1 | 1 |
| 6 | 1 | 1 | 0 | 1 |
| 7 | 1 | 1 | 1 | 1 |

## A. Test Generation

We start with an X test cube for all C-testable circuit's primary inputs. For test generation, we select an arbitrary cell and perform row justification and row propagation subsequently, until all-X test cube evolves into a complete test vector. Fig. 3 shows all TPG steps for our 8-bit RCA example.

The procedure starts by a test selected from the simple row propagation group (row $S_1$, Fig. 3) (i.e., 0, 2, 3, 4, and 5), and the arbitrary cell test pattern (test 0) is applied to an arbitrary cell (row $S_2$, Fig. 3). For row justification, $C_{in}$ values can be justified by any test with $C_{out}$ identical to the original cell's $C_{in}$. However, to achieve minimum test vector count, best is to select the same cell test pattern (test 0) in row justification process (row $S_3$, Fig. 3). In row propagation (selected cell's left-side neighbors, Fig. 2), a cell test pattern is needed with $C_{in}$ identical to the original cell's $C_{out}$. Hence, the identical cell test pattern (test 0) is selected for row propagation (row $S_4$, Fig. 3). This procedure continues until all cell test patterns of a simple row propagation group (i.e., 0, 2, 3, 4, 5, and 7) are applied to the arbitrary cell.

| Simple Row Propagation Group | | | | | | | |
|---|---|---|---|---|---|---|---|
| Num | A7B7 | A6B6 | A5B5 | A4B4 | A3B3 | A2B2 | A1B1 | A0B0 |
| S1 | XX | XX | XX | XX | XX | XX | XX | XX |
| S2 | XX | XX | XX | XX | 00 | XX | XX | XX |
| S3 | XX | XX | XX | XX | 00 | 00 | 00 | 00 |
| S4 | 00 | 00 | 00 | 00 | 00 | 00 | 00 | 00 |
| Complementary Row Propagation Group | | | | | | | |
| Num | A7B7 | A6B6 | A5B5 | A4B4 | A3B3 | A2B2 | A1B1 | A0B0 |
| C1 | XX | XX | XX | XX | XX | XX | XX | XX |
| C2 | XX | XX | XX | XX | 00 | XX | XX | XX |
| C3 | XX | XX | XX | XX | 00 | 11 | 00 | 11 |
| C4 | 00 | 11 | 00 | 11 | 00 | 11 | 00 | 11 |
| C5 | 11 | 00 | 11 | 00 | 11 | 00 | 11 | 00 |

Fig. 3. TPG for simple and complementary row propagation groups

After applying all six simple row propagation cell test patterns, we start fresh with an all-X test cube to start complementary row propagation group (row $C_1$, Fig. 3). Then, a cell test pattern from the complementary row propagation group (e.g., cell test pattern 1) is selected (row $C_2$ Fig. 3). For row justification, the lower order cell must be tested by a cell test pattern whose $C_{out}$ response is identical to the chosen cell $C_{in}$. Since, in this group, $C_{in}$ is different from $C_{out}$, the identical cell test pattern cannot be selected. Instead, we select a cell test pattern within the same group with complementary $C_{in}$ and $C_{out}$. To have minimum test vector count, an unused cell test pattern is preferred. Hence, cell test pattern 6 is selected for the immediate lower order cell of the original cell. By repeating the same procedure, we toggle between the first and the second cell test patterns until we reach the RCA's right-most cell meaning that alternating cell test patterns 1 and 6 can be applied to all right-side neighbors of the original cell (row $C_3$, Fig. 3).

For row propagation, a similar procedure is followed. This makes all left-side neighbors of the original cell receive the same alternating cell test patterns. In our example, test 6 and test 1 will be repeated for the rest of the cells. This process results in a test vector for the C-testable circuit consisting of two alternating cell test patterns (row $C_4$, Fig. 3). This process is completed upon addition of another test vector reversing the alternating cell test patterns (row C5 Fig. 3).

Abovementioned process is repeated for simple and complementary row propagation groups until all propagation table rows are marked. In our example, rows 2, 3, 4, 5, and 7 complete simple row propagations, resulting in 5 more test vectors.

This section have discussed the C-testability of RCA whose cells are chained in only one direction. Shown in Fig. 4, in an array MULT, two chains (row and column) exist making test generation process more complex. Hence, based on this section, we extend our method for an array MULT. Unlike RCA, the array MULT is not C-testable; however, its C-testable test generation process is discussed based on its bijectivity.

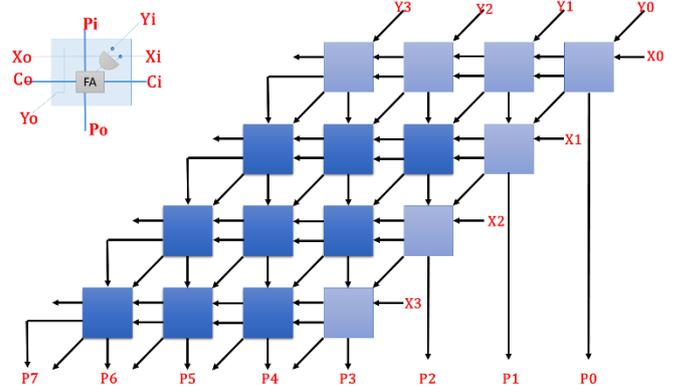

Fig. 4. Array multiplier

## V. MULT C-TESTABILITY

An array MULT is a 2-D array of cells, each cell has an AND gate and a FA, with two n-bit vector inputs and a 2n-bit vector output. The array MULT's propagation table (Fig. 5(d)) lists the array MULT cell's parallel inputs, chain inputs and outputs. The cell's parallel inputs $X_{in}$ and $Y_{in}$ are not used independently, so, they are ANDed $X.Y_{in}$ (Fig. 5(d)'s input list). Justification of values to $X_{in}$ and $Y_{in}$ will be discussed in test generation subsection (Subsection A). Considering $XY_{in}$ as cell's parallel input, the MULT cell chain inputs are $C_{in}$ and $P_{in}$, and cell's chain outputs are $C_{out}$ and $P_{out}$, respectively. The array MULT cells' $C_{in}/C_{out}$ are chained in rows (*row chain*) and their $P_{in}/P_{out}$ are chained in columns (*column chain*).

For each cell, a propagation table's row is selected and applied as discussed in Section IV. $XY_{in}$, is directly determined by the table's selected row. However, $C_{in}$ and $P_{in}$ inputs must be justified by back-tracing to array MULT's borders. $C_{in}$ is justified by *row justification* and the $P_{in}$ is justified by *column justification*. Also, *row propagation* and *column propagation* must be done to propagate the obtained primary inputs to the entire circuit in row and column chains, respectively.

In justification of a cell's chain input values, three test groups exist: (*i*) tests for which both row and column chain inputs are the same as the chain outputs (*simple propagation*); (*ii*) tests having similar input and output in the row chain ($C_{in}/C_{out}$) and complementary values for the column chain input and output ($P_{in}/P_{out}$)(*complementary column propagation);* and (*iii*) tests having complementary values of input and output in both row and column chains (*total complementary propagation)*. Test generation is performed based on this cell test pattern grouping. Using the aforementioned procedure and propagation table, our test generation process is discussed in next subsection.

## A. Test Generation

Similar to the C-testable adder, cell test patterns are selected based on cell test pattern grouping and then justification and propagation are done subsequently.

For justification and propagation, we select cell test patterns enabling justification and propagation chosen from the same group that the original cell test pattern has previously been selected. For this example, cell test pattern groups based on same or complementary chain values are shown in Fig. 5(d).

We start with a test from the simple propagation group (i.e., test 0, 1, 6, 7); a cell test pattern (e.g. test 0) is applied to an arbitrary cell (similar to adder TPG in Section IV). For row justification, justification of the $C_{in}$ value can be done by any test whose $C_{out}$ is identical to the original cell's $C_{in}$. As discussed in Section IV, the same cell test pattern (test 0) is selected for row justification. Similarly, column justification of the $P_{in}$ value of our arbitrary cell is done by the same cell test pattern (test 0). Fig. 5(a) shows how P values propagate vertically as same P values in the upper cells.

For row justification, every cell visited, causes column justification until MULT's primary inputs are reached. Alternatively, for column justification, every cell visited, causes row justification until primary inputs are reached. Row and column propagation is done using the same cell test pattern until primary outputs are reached. This process continues until all test patterns of the simple propagation group (i.e., 0, 1, 6, 7) are applied to the arbitrary cell.

After all four simple propagation cell test patterns are applied, a test (e.g., test 3) from complementary column propagation group (e.g., test 3 and 4) will be selected. For row justification, the same cell test pattern (test 3) is selected, due to identical $C_{in}$ and $C_{out}$. However, for column justification, the lower (vertical) order cell must be tested by a cell test pattern with $P_{out}$ identical to arbitrary cell's $P_{in}$. Since in this group (tests 3 and 4), $P_{in}$ is different from $P_{out}$, the same cell test pattern cannot be selected. Instead, we select a cell test pattern from the same group (same $C_{in}$, $C_{out}$) with $P_{in}$ and $P_{out}$ complements. Considering abovementioned conditions, cell test pattern 4 is selected for the original cell's immediate upper-level cell (Fig. 5(b)). Repeating the same procedure for column justification, we toggle between the first and the second tests (3 and 4) until MULT's uppermost cell is reached (primary inputs). Similar procedure is repeated for row justification of the $C_{in}$ values of the original cell's upper cells. This procedure results in applying cell test patterns 3 and 4 to the upper cells of the original cell, with the same cell test pattern to the right of each cell.

In row and column propagation procedures for the original cell and its upper- and right-side neighbor cells, a similar procedure is followed. This procedure makes all rows on the upper and lower sides of the original cell receive the same alternating test patterns (i.e., test patterns 3 and 4). This process results in a test vector for the C-testable circuit consisting of two alternating cell test patterns. We complete this process by adding another test vector that reverses the alternating cell test patterns, i.e., test vector 4 and 3 in this group.

As a final step of the test generation, a cell test pattern from the totally complementary propagation group (i.e., test 2 and 5) is selected (e.g., test 2). A similar procedure used for cell test patterns 3 and 4 is repeated (2 and 5). This procedure results in alternating cell test patterns in a checker board pattern for the entire array MULT (Fig. 5(c)). Fig. 5(c) shows how P values propagate vertically to be P complements, and how C values propagate horizontally to be C complements. As before, this process is completed by reversing cell test patterns to 5, and 2.

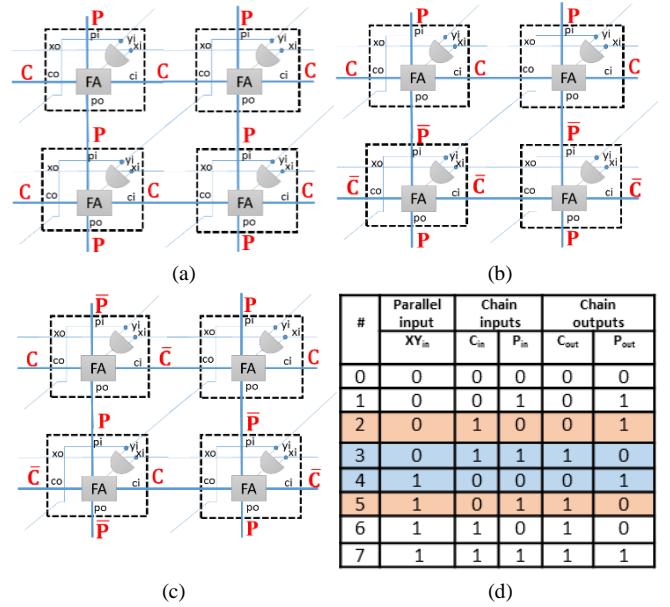

Fig. 5. (a) Simple propagation group, (b) Complementary column propagation group, (c) Totally complementary propagation group, (d) propagation table for MULT cell

The above process for simple, complementary column and totally complementary propagation groups will be repeated for all propagation table's rows. In our example, rows 1, 6, and 7 must be considered to complete simple propagations, resulting in three more test vectors.

The aforementioned procedure determines cell tests for $C_{in}$, $P_{in}$, and $XY_{in}$ shown in Fig. 5(d). We assumed independent X and Y inputs so far, which is incorrect for cells in a row with the same X and in a column with the same Y. Hence, we need to justify XY (AND of X and Y) values on the X and Y primary inputs. For the simple propagation group where all X and Y values are the same, justification is done by assigning appropriate values to X and Y inputs. For example, justifying XY value of 0 demands setting either all X primary input values to 0 or all Y primary input values to 0. For those test vectors requiring alternating XY values for the neighboring cells, values 0 and 1 must be properly interlaced on X and Y primary input bits to satisfy the required cell values.

### B. MULT's Borderlines

Our TPG procedure explained in Subsection A fails once reaches the MULT's left-side border where $C_{out}$ becomes product inputs of the succeeding lower cells. Considering right- and left-side border P inputs and outputs, we observe that all $P_{out}$s are used for justifications, except the right-side border $P_{out}$s that are left open and unused. Additionally, all $P_{in}$s except the left-side border $P_{in}$s justified toward the P primary input vector. Left-side border $P_{in}$s are not justified at all. We can achieve both goals using a multiplexer connecting these two when MULT is in the test mode. In the test mode, multiplexers shown in Fig. 6 (left side) find ways for side $P_{in}$ values to be justified through the otherwise unused $P_{out}$s on the right-side border.

## VI. COMBINED WITH CELL BASED DETERMINISTIC TPG

C-Testability related methods have limited themselves to use exhaustive test generation for each cell. However, because of redundant faults, exhaustive testing does not need to achieve 100% fault coverage. Instead, deterministic test generation can be used for each cell. In other words, using a closed set of cell test pattern is a vital condition for C-testability. It means that by applying each cell test pattern of the set to an arbitrary cell, the propagation and justification procedures do not produce a cell test pattern out of the set for other cells.

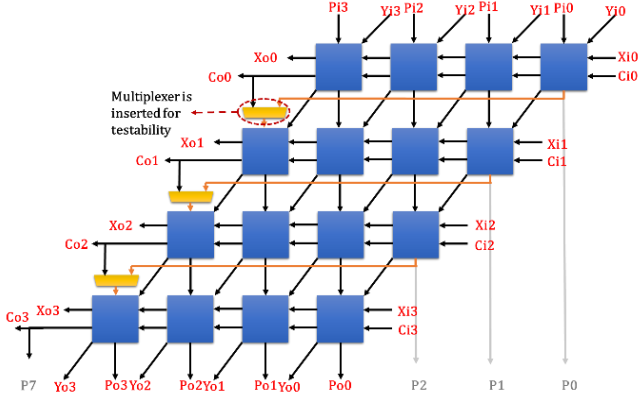

Fig. 6. Array multiplier design for test using C-testability

### A. Adder Cell Deterministic Testing

The gate-level adder cell (FA) and its reduced faults using fault collapsing are shown in Fig. 7(a). Using deterministic TPG, a test set is generated with five cell test patterns covering all the faults. Referring to Table I, cell test patterns 1, 2, 3, 4, and 6 alone cover all FA's faults. Using XOR gates for the sum output reduces test vector count to only four (i.e., 1, 3, 4 and 6).

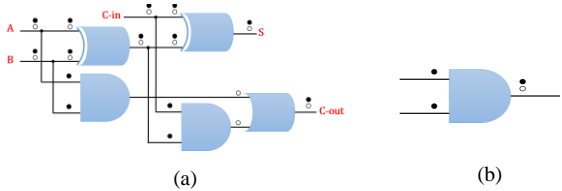

Fig. 7. (a) FA cell, (b) AND cell reduced faults for deterministic TPG

### B. MULT Cell Deterministic Testing

The MULT cell is similar to an adder cell except that one of its inputs is the AND gate's output. Shown in Fig. 7(b), only two faults are added to the FA cell faults covered by ($X_{in}$, $Y_{in}$) = (1, 0) and (0, 1). When FA and the AND gate test patterns are combined, the test vector count per cell for a MULT cell will be five. The two test patterns for the AND gate must be covered by the MULT cell test patterns. Test patterns for a MULT cell ($X_{in}$, $Y_{in}$, $C_{in}$, $P_{in}$) are as follows: (1,0,0,1), (1,0,1,0), (0,1,1,1), (1,1,0,0), (1,1,0,1). The first test pattern can be omitted if XOR gates are used for the sum output.

The MULT test vectors and their responses (primary I/Os) are shown in Table II. Each of the values must be repeated to cover all primary inputs or outputs. An even number of bits for the MULT operands is assumed. Single values in the table apply uniformly to all MULT bits. On the other hand, multiple values are for a pair of adjacent MULT cells.

## VII. TPG AND ORA

Our proposed test method can be applied to a design in various ways. Most of them like BIST and boundary scan do not impose any overheads on the pin count. A BIST architecture for this purpose requires a 3-bit counter to count for five test vectors and a decoder to produce iterative input test patterns and their responses. The twelve-bit output of the 3-input decoder provides inputs and outputs of two adjacent MULT cells. Twelve outputs are composed of eight-bit test vectors for test pattern generator and four-bit response patterns for Output Response Analyzer (ORA). The decoder is designed through solving multiple output minimization using Karnaugh Map (K-map) for twelve outputs and three inputs. The hardware, power and delay of our pattern generator are 16.28%, 27.58% and 37.5% respectively less than those of a design employing multiplexer to set the patterns.

TABLE II. ALL 5 TEST VECTORS AND RESPONSES, VALUES ARE FOR CELL/**ADJACENT CELL**

| Inputs | | | | Outputs | |
|---|---|---|---|---|---|
| $X_{in}$ | $Y_{in}$ | $C_{in}$ | $P_{in}$ | $C_{out}$ | $P_{out}$ |
| 1 | 0 | 0 | 1 | 0 | 1 |
| 0/1 | 1 | 1/**0** | 0 | 1/**0** | 0 |
| 1/0 | 1 | 0/**1** | 1 | 0/**1** | 1 |
| 1 | 1/**0** | 1 | 1/**0** | 1 | 0/**1** |
| 1 | 0/**1** | 0 | 0/**1** | 0 | 1/**0** |

Fig. 8 shows the flow and organization of our proposed MULT's BIST architecture. The test pattern generator applies tests to the MULT in the test mode in five clock cycles and then compares outputs against the golden expected results.

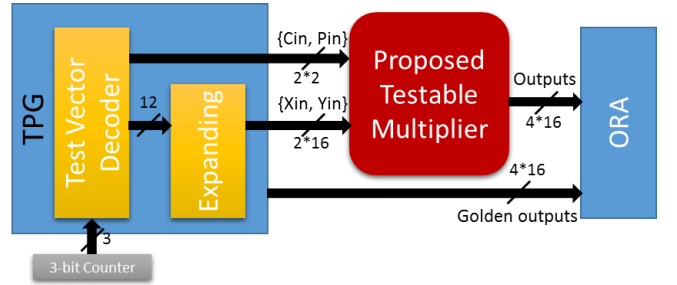

Fig. 8. BIST design for array multiplier

## VIII. EXPERIMENTAL RESULTS

Using fault simulation, we have verified that five test vectors (Table II) cover all MULT's stuck-at faults when in test mode.

Performance results for a 16-bit MULT, including area, delay and power are obtained after synthesis in 90 nm technology. The basic MULT, the MULT with multiplexers used for test-mode (DFT), and the MULT with inserted BIST are compared as presented in Table III. The TPG and ORA used for the BIST are also separately evaluated. Area Overhead (AO), Delay Overhead (DO) and Power Overhead (PO) are reported for each case. Shown in Table III, BIST components, TPG and ORA impose 8.29% and the DFT modification imposes 1.57% area overhead over the basic MULT.

### A. Comparisons

Table IV compares the result of our proposed method against those of other existing methods based on C-testability. Our proposed design can achieve 100% Fault Coverage (FC) with a very small and fixed text vector count (TVC) of 5 independent of the MULT size. Both small TVC and an efficient BIST design result in low Test Time (TT).

TABLE III. Experimental Results

| Design | Overheads | | | | | |
|---|---|---|---|---|---|---|
| | Area ($\mu m^2$) | AO% | Delay (ns) | DO% | Power ($\mu w$) | PO% |
| *Basic MULT* | 6729 | - | 5.10 | - | 509.8 | - |
| *DFT MULT* | 6834 | **1.57** | 5.12 | 0.39 | 514.5 | 0.92 |
| *BIST+DFT MULT* | 7600 | 13.00 | 5.22 | 2.35 | 547.8 | 7.46 |
| *TPG+ORA* | 557 | **8.29** | 0.51 | - | 25.92 | 4.03 |

A DFT method for MULT testing was presented in [6]. Both area overhead and test vector count of our method for a 16-bit MULT are less than those of the method presented in [6]. The method of [7] imposes no DFT hardware overhead; however, our BIST area overhead (including the DFT overhead) is less than that of [7]. Authors of [7] have compared their method against a deterministic (DET) test pattern generator. Neither the method of [7] nor the DET of [7] can achieve 100% FC. However, our testable design, achieving 100% FC, requires less test vectors compared to the method of [7] and DET of [7]. Obviously, the reported FC does not include the DFT hardware.

TABLE IV. Comparison of Design with Other Methods

| Design | TVC | TT (clk) | FC% | AO% |
|---|---|---|---|---|
| *Proposed DFT MULT* | 5 | - | 100 | 1.57 |
| *Proposed BIST+DFT MULT* | 5 | 6 | 100 | 13.00 |
| *[6]* | 74 | - | 100 | 4.17 |
| *[7] (BIST)* | 165 | 177 | 98.9 | 14.9 |
| *[7] (DET)* | 59 | - | 99.98 | - |

### B. Scalability Issue

Our method proposes a better scalability when compared with state-of-the-art as detailed in Table V. Area, delay and power overheads of our proposed method are decreased by increasing the MULT size. For example, for a 64-bit MULT, our method's overheads are less than 0.5%, while [6] has fixed overheads for any MULT size. The test vector count for the method presented in [7] increases linearly, while our proposed method maintains fixed number of test vectors for different array MULT sizes. Table V demonstrates the changes occurring because of array MULT size increasing.

TABLE V. Scalability Comparison of Various Designs

| Design | Overheads | TVC | FC% |
|---|---|---|---|
| *Proposed DFT MULT* | decreasing | fixed | 100 |
| *[6]* | fixed | fixed | 100 |
| *[7] (without BIST)* | 0 | Increasing linearly | <100 |
| *[7] (DET)* | 0 | increasing | <100 |

## IX. CONCLUSION

In this paper, a method is proposed for testing array multipliers. The main concept used is C-testability, and the multiplier is modified so that C-testability can be used for generating tests. The modifications to the original multiplier for making it C-testable are considered as a DFT method for the multiplier. This DFT has a minimal area, delay, and power overhead. The result of this work is a very fast multiplier testing with only five test vectors.